\documentclass[12pt]{article}

\usepackage{amsfonts,amscd,amssymb,amsmath}
\usepackage{verbatim}

\date{}

\begin{document}

\title{On  Measurements, Numbers and $p$-Adic Mathematical Physics}

\author{{Branko Dragovich}
\\ \small{Institute of Physics, University of Belgrade, Pregrevica 118, 11080 Belgrade, Serbia }\\
 \, dragovich@ipb.ac.rs  }

\maketitle
\begin{abstract}
In this short paper I consider relation between  measurements,
numbers and $p$-adic mathematical physics. $p$-Adic numbers are
not result of measurements, but nevertheless they play significant
role in description of some systems and phenomena. We illustrate
their ability for applications referring to some sectors of
$p$-adic mathematical physics and related topics, in particular,
to string theory and the genetic code.

\end{abstract}

\hskip 2.3 truecm {\it Dedicated to Igor V. Volovich on the
occasion of his 65th birthday}

\section{Introduction}

It is well known that science is based on experiments and their
theoretical modelling.
 To my knowledge, Igor Volovich was the first
mathematical physicist who pointed out the field $\mathbb{Q}$ of
rational numbers as a bridge between  experiments and their
possible theoretical descriptions \cite{volovich0}. Namely, the
results of experimental measurements are  some rational numbers,
presented by finite number of digits. From mathematical point of
view, $\mathbb{Q}$ is dense in the field $\mathbb{R}$ of real
numbers and the fields $\mathbb{Q}_p$ of $p$-adic numbers, for all
primes $p$. On $\mathbb{R}$ and $\mathbb{Q}_p ,$ as well as on
their algebraic extensions, there are well developed analysis.
Analysis on $\mathbb{R},$ and the field $\mathbb{C}$ of complex
numbers, is well employed in description of physical systems. In
1987 \, I. Volovich proposed \cite{volovich1} to use $p$-adic
numbers in description of space-time at the Planck scale and in
string theory, and he introduced concept of $p$-adic strings. It
was first valuable application of $p$-adic analysis in
mathematical physics and was marked beginning of $p$-adic
mathematical physics (for reviews see
\cite{freund,vladimirov,dragovich}). I am very glad for
opportunity to collaborate with I. Volovich on application of
$p$-adic numbers and adeles in various sectors of $p$-adic
mathematical physics from its very beginning.

In the sequel of this paper I shall discuss some aspects of
measurements, $p$-adic numbers and adeles, and their role in
$p$-adic mathematical physics and related topics. In particular, I
want to point out that results of measurements are rational
numbers endowed by Archimedean norm and  natural ordering. Also I
want to emphasize that $p$-adic numbers, although being not direct
result of measurements, are very important tools in description of
some systems and phenomena.  I shall demonstrate it at two
examples -- string theory and the genetic code.

\section{Measurements, Numbers and Their Applicability}

Any measurement  is comparison of two quantities of the same kind,
one quantity by convention  is taken to be unit of measurement and
the other one is subject of measuring.  Comparing these two
quantities we get their ratio, which contains some information how
many times measured quantity is larger or smaller than the unit
quantity. This ratio is a rational number, with only some finite
number  of  certain  digits. More precise measuring usually
results in more certain digits. However, due to many reasons,
there is always some uncertainty and result of measurement has
only an approximative character. It is worth noting that these
digits of the obtained rational number are digits in the decimal
expansion of the real number. Hence, results of measurements are
rational numbers with Archimedean norm and natural ordering, which
are characteristics of real but not of $p$-adic numbers. Although
this assertion is evident, it is not generally recognized.

Recall that, from algebraic point of view, $\mathbb{Q}$ is a field
and any its element $x$ can be presented in the form $x =
\frac{m}{n},$ where $m \in \mathbb{Z}$ and $n \in \mathbb{N}.$ It
is well known that summation and multiplication, and their inverse
operations, are well defined for these numbers. Among rational
numbers there is also another property, which is distance and is
related to the norm. According to the Ostrowski theorem
\cite{schikhof} there are no other non-trivial norms on
$\mathbb{Q},$ which are non-equivalent either to the absolute
value (Archimedean)  or $p$-adic (non-Archimedean) norm,  related
to prime numbers. A rational number $x = \frac{m}{n} = p^\nu \,
\frac{a}{b}$, where integers $a$ and $b \neq 0$ are not divisible
by prime number $p$, by definition has $p$-adic norm $|x|_p =
p^{-\nu}$ and $|0|_p = 0 .$ Since $|x + y|_p \leq \mbox{max}
\{|x|_p \,, |y|_p \}$, $p$-adic norm is a non-Archimedean
(ultrametric) one. As completion of $\mathbb{Q}$ with respect to
the absolute value $|\cdot|_\infty$ gives the field
$\mathbb{Q}_\infty \equiv \mathbb{R}$ of real numbers, by the same
procedure using $p$-adic norm $|\cdot|_p$ one gets the field
$\mathbb{Q}_p\,\,$ of $p$-adic numbers (for any prime number $p =
2,\, 3\,, 5\, \cdots$). Any number $0 \neq x \in \mathbb{Q}_p\,\,$
has its unique canonical representation (see, e. g.
\cite{vladimirov})
\begin{align} \label{equ.1}
x = p^{\nu} \, \sum_{n = 0}^{+ \infty} \, x_n\, p^n  \,, \quad \nu
 \in \mathbb{Z}\,, \quad x_n \in \{0,\, 1,\, \cdots, \, p-1 \},
\quad x_0 \neq  0 .
\end{align}
From representation (\ref{equ.1}) one can conclude that rational
number obtained in the process of measuring cannot  have this form
with $p$-adic (non-Archimedean) norm. Also, $p$-adic numbers have
not natural ordering, while results of measuring have it.

Measurements  often are not direct, but with help of some tools
and can be viewed as measuring of a length. By this way
measurement is related to the Archimedean axiom in geometry, which
is originally formulated for two segments on a straight line and
it states:
 a larger segment $(A)$ can be always surpassed by some finite number
$(n)$ of the successive addition of  a smaller segment $(a)$ along
the larger one. This can be also expressed in terms of two  real
numbers: Let $a, A \in \mathbb{R}$ and  $ 0<|a|_\infty <
|A|_\infty$ then there is always $n \in \mathbb{N}$ such that
$|n|_\infty |a|_\infty > |A|_\infty.$  Note that addition of a
smaller segment along the larger one is just as measuring the
larger segment by the smaller one. In the $p$-adic case,
Archimedean axiom is not valid, because $|n|_p \leq 1.$ Recall
that one cannot measure distances smaller than the Planck length
$\ell_P = \sqrt{\frac{\hbar G}{c^3}} \sim 10^{-35} m$, because
quantum and gravity effects lead to the uncertainty which cannot
be smaller than the Planck length $\ell_P$. This was motivation
that I. Volovich conjectured \cite{volovich0,volovich1} existence
of non-Archimedean geometry with $p$-adic numbers at the Planck
scale.

From achievements  of applied mathematics it follows that systems
of real numbers which have field structure, i.e. $\mathbb{R}$ and
$\mathbb{C}$, have the most applicability. For example, classical
and quantum theoretical physics are mainly based on analysis
related to maps $\mathbb{R}  \to \mathbb{R}$ and $\mathbb{R} \to
\mathbb{C}, $ respectively. Comparing to $\mathbb{R}$ and
$\mathbb{C}$,  normed division algebra of quaternions
$\textbf{H},$ which is noncommutative, is rather less applicative.
Algebra of octonions $\textbf{O}$ as the last normed division
algebra,  which is noncommutative and nonassociative, has a minor
role in applications. It is worth noting that complex numbers,
which are not result of direct measuring and  are unordered, are
unavoidable in quantum mechanics, where they are used not for
description of space-time but for complex-valued wave functions,
which contain all information about the state of the quantum
system.

Now one can pose the following question: Being not results of
measurements, what role $p$-adic numbers can play in description
of physical or some other systems? $p$-Adic numbers should play
unavoidable role where description with real numbers, or with
systems of real numbers, is inadequate. In physical case such
situation should be at the Planck scale, because it is not
possible to measure distances smaller than the Planck length.  It
should be also the case with very complex physical, living,
cognitive, information and some other complex systems and
phenomena. For example, such systems may have space of states with
some $p$-adic structure, like it is the case with quantum states
described by complex-valued wave functions. Thus, we expect
inevitability of $p$-adic numbers at some more profound levels in
understanding  of the Universe in its parts as well as a whole
\cite{arefeva,dragovich1,dragovich2}. The first step towards
possible $p$-adic level of knowledge is invention of relevant
mathematical tools and construction of the adequate physical
models. This is subject of $p$-adic mathematical physics and its
brief recent overview is presented in \cite{dragovich} (see also
\cite{dragovich3}).


\section{$p$-Adic Numbers in Applications}

$p$-Adic numbers are discovered by Kurt Hansel in 1897 as a new
tool in number theory. They have applications in many parts of
modern mathematics: number theory, algebraic geometry, theory of
representations, ...
 $\mathbb{Q}_p$ is locally compact, complete and totally
disconnected topological space. There is  rich structure of
algebraic extensions of $\mathbb{Q}_p$.

There are many possibilities for mappings between $\mathbb{Q}_p$.
The most elaborated is analysis related to mappings $\mathbb{Q}_p
 \to \mathbb{Q}_p$ and $\mathbb{Q}_p \to \mathbb{C}$. Usual complex valued
 functions of $p$-adic argument are additive $\chi_p(x) =  e^{2\pi i \{x \}_p}$ and multiplicative
 $|x|^s$ characters, where $\{x \}_p$ is fractional part of $x$ and $s \in \mathbb{C}$ (for many aspects of
 $p$-adic numbers and their analysis, we refer to \cite{freund,vladimirov,schikhof,gelfand}).

Real and $p$-adic numbers, as completions of $\mathbb{Q},$ are
joined into adeles (see, e. g.
\cite{freund,vladimirov,gelfand,weil}). An adele $\alpha$ is an
infinite sequence made od real and $p$-adic numbers in the form
\begin{align} \label{equ.3.1}
\alpha = (\alpha_\infty ,\, \alpha_2 ,\, \alpha_3, \, \cdots ,\,
\alpha_p \,,\, \cdots) \,,           \quad \alpha_\infty \in
\mathbb{R} \,, \,\, \alpha_p \in \mathbb{Q}_p \,,
\end{align}
where for all but a finite set $\mathcal{P}$ of primes $p$ it has
to be $\alpha_p \in \mathbb{Z}_p = \{ x\in \mathbb{Q}_p \, : |x|_p
\leq 1 \}$.  $\mathbb{Z}_p$ is the ring of $p$-adic integers,
which they have $\nu \geq 0$ in (1). The set
$\mathbb{A}_{\mathbb{Q}}$ of all  adeles can be presented as
\begin{align} \label{equ.3.2}
\mathbb{A}_{\mathbb{Q}} = \bigcup_{\mathcal{P}} A (\mathcal{P})\,,
\quad A (\mathcal{P}) = \mathbb{R}\times \prod_{p\in \mathcal{P}}
\mathbb{Q}_p \times \prod_{p \not\in \mathcal{P}} \mathbb{Z}_p \,.
 \end{align} Elements of $\mathbb{A}_{\mathbb{Q}}$ satisfy componentwise addition and
multiplication and form the adele ring. Adeles and their functions
are useful for connection of real and $p$-adic models of the same
system, see, e.g.
\cite{freund,vladimirov,dragovich,dragovich3,dragovich4}.

 In the sequel we
shall consider $p$-adic strings and $p$-adic structure of the
genetic code.

\subsection{$p$-Adic Strings}

 $p$-Adic strings are introduced by
construction of their scattering amplitudes in analogy with
ordinary strings \cite{volovich1,freund1}. The simplest amplitude
is for scattering of two open scalar strings. The
 crossing symmetric Veneziano amplitude for ordinary strings is
\begin{align} \label{equ.2}
 A_\infty (a, b) = g_\infty^2 \,\int_{\mathbb{R}}
|x|_\infty^{a-1}\, |1 -x|_\infty^{b-1}\, d_\infty x   = g_\infty^2
\, \frac{\zeta (1 - a)}{\zeta (a)}\, \frac{\zeta (1 - b)}{\zeta
(b)}\, \frac{\zeta (1 - c)}{\zeta (c)},
\end{align}
where $a+b+c = 1.$ The crossing symmetric Veneziano amplitude for
scattering of two open scalar $p$-adic strings is direct analog of
(\ref{equ.2}) \cite{freund1}, i. e.
\begin{align} \label{equ.3}
A_p (a, b) = g_p^2 \, \int_{\mathbb{Q}_p} |x|_p^{a-1}\, |1
-x|_p^{b-1}\, d_p x   = g_p^2 \, \frac{1 - p^{a - 1}}{1 -
p^{-a}}\, \frac{1 - p^{b - 1}}{1 - p^{-b}}\, \frac{1 - p^{c -
1}}{1 - p^{-c}}.
\end{align}
Integral expressions in (\ref{equ.2}) and (\ref{equ.3}) are the
Gel'fand-Graev-Tate beta functions on $\mathbb{R}$ and
$\mathbb{Q}_p$, respectively \cite{gelfand}. Note that  ordinary
and $p$-adic strings differ only in description of their
world-sheets -- world-sheet of $p$-adic strings is presented by
$p$-adic numbers.  Kinematical variables contained in $a, b, c$
are the same real (complex) numbers in both cases. Veneziano
amplitude for $p$-adic strings (\ref{equ.3}) is rather simple and
presented by elementary functions.

The above Veneziano string amplitudes are connected by the
Freund-Witten product formula \cite{freund2}:
\begin{align} \label{equ.3.7}
   A (a, b) =
 A_\infty (a, b) \prod_p A_p (a, b) = g_\infty^2 \,\prod_p g_p^2 = const.
\end{align}
 Formula (\ref{equ.3.7}) follows from the Euler
product formula for the Riemann zeta function applied to $p$-adic
string amplitudes (\ref{equ.3}). Main significance of
(\ref{equ.3.7}) is in the fact that scattering amplitude for real
string  $A_\infty (a, b)$, which is a special function, can be
presented as product of inverse $p$-adic amplitudes, which are
elementary functions. Also, this product formula treats $p$-adic
and ordinary strings at the equal footing. It gives rise to
suppose that if there exists an ordinary scalar string then it
should exist also its $p$-adic analog.

It is remarkable that there is an effective field theory
description of the above  open  $p$-adic strings. The
corresponding Lagrangian is very simple and  at the tree level
describes not only four-point scattering amplitude but also all
higher  ones. The exact form of this  Lagrangian for effective
scalar field $\varphi,$ which describes open $p$-adic string
tachyon, is \cite{freund3,frampton}
\begin{align} \label{equ.4}
 {\cal L}_p = \frac{m^D}{g^2}\, \frac{p^2}{p-1} \Big[
-\frac{1}{2}\, \varphi \, p^{-\frac{\Box}{2 m^2}} \, \varphi +
\frac{1}{p+1}\, \varphi^{p+1} \Big],
\end{align}
where $p$
 is any prime number, $D$ -- space-time dimensionality, $\Box = - \partial_t^2  + \nabla^2$ is the
$D$-dimensional d'Alembertian and  metric has signature $(- \, +
\, ...\, +).$ This is nonlocal and nonlinear Lagrangian.
Nonlocality is in the form of infinite number of space-time
derivatives
\begin{align} \label{equ.5}
  p^{-\frac{\Box}{2 m^2}} = \exp{\Big( - \frac{\ln{p}}{2 m^2}\,
\Box \Big)} = \sum_{k \geq 0} \, \Big(-\frac{\ln p}{2 m^2} \Big)^k
\, \frac{1}{k !}\, \Box^k
\end{align}
and it is a consequence of  strings as extended objects.

It is worth noting that Lagrangian (\ref{equ.4})  can be rewritten
in the  form \cite{dragovich4}
\begin{align}
 {\cal L}_p = \frac{m^D}{g^2}\, \frac{p^2}{p-1} \Big[
\frac{1}{2}\, \varphi \, \int_{\mathbb{R}}
\Big(\int_{\mathbb{Q}_p\setminus \mathbb{Z}_p} \chi_p (u)
|u|_p^{\frac{k^2}{2m^2}} du \Big) \tilde{\varphi}(k)\, \chi_\infty
(kx)\, d^4k  + \frac{1}{p+1}\, \varphi^{p+1} \Big], \label{equ.10}
\end{align}
where $\chi_\infty (kx) = e^{-2\pi i kx}$ is the real additive
character. Since $\int_{\mathbb{Q}_p} \chi_p(u) |u|^{s-1} du =
\frac{1 - p^{s-1}}{1-p^{-s}} = \Gamma_p (s)$ and it is present in
the scattering amplitude (\ref{equ.3}), one can think that
expression $\int_{\mathbb{Q}_p\setminus \mathbb{Z}_p} \chi_p (u)
|u|_p^{\frac{k^2}{2m^2}} du$ in (\ref{equ.10}) is related to the
$p$-adic string world-sheet.

Now suppose that  Lagrangian (\ref{equ.4}), which is written only
in terms of  real numbers, and its scattering amplitude
\begin{align} \label{equ.11}
A_p (a, b)  = g_p^2 \, \frac{1 - p^{a - 1}}{1 - p^{-a}}\, \frac{1
- p^{b - 1}}{1 - p^{-b}}\, \frac{1 - p^{c - 1}}{1 - p^{-c}}
\end{align}
were discovered first and after that it was found integral
representation (\ref{equ.3}) to scattering amplitudes. Then it
would be natural to  conclude that Lagrangian (\ref{equ.4})
describes quite new strings which have $p$-adic world-sheet and
can be presented in the equivalent form (\ref{equ.10}). By this
way, with help of $p$-adic numbers we are able to get some more
profound information about structure of the system  which is not
accessible to direct measuring. For some other examples of
physical models, which also include both $p$-adic valued and real
(complex) valued functions of $p$-adic argument, we refer to
\cite{dragovich} and references therein.

\subsection{$p$-Adic Structure of  the Genetic Code}

The genetic code is connection between $64$ codons, which are
building blocks of the genes, and $20$ amino acids, which are
building blocks of the proteins. In addition to coding amino
acids, a few codons code stop signal, which is at the end of genes
and terminates process of protein synthesis.  Codons are ordered
triples composed of C, A, U (T) and G nucleotides. Each codon
presents an information which controls use of one of the 20
standard amino acids or stop signal in synthesis of proteins. It
is obvious that there are $4 \times 4\times 4 = 64$ codons. For
molecular biology and the genetic code, one can see, e.g.
\cite{watson}.

From mathematical point of view, the genetic code is a mapping of
a set of $64$ elements onto a set of $21$ elements. There is in
principle a huge number of possible mappings, but the genetic code
is one definite mapping with a few slight modifications. Hence the
main problem is to find structure of $64$ elements which is used
in mapping that corresponds to the genetic code.
 It will be demonstrated here that the set of $64$ codons has $p$-adic
structure, where $p = 5$ and $2$. Detail exposition of $p$-adic
approach to the genetic code is presented in
\cite{dragovich5,dragovich6,dragovich7} (see also \cite{kozyrev}
for a similar consideration).

 The idea behind  this approach is as follows. Codons
which code the same amino acid should be in information sense
close. To quantify this closeness (nearness) we should use some
distance. Ordinary distance is appropriate to determine spatial
distribution of codons but not for their distribution with respect
to information characteristics. From insight to the table of the
genetic code (see, e. g. Table 1) one can conclude that
distribution of codons is like an ultrametric tree and it suggests
to use $p$-adic distance.

To this end, let us  introduce the following subset  of natural
numbers:
\begin{equation} \mathcal{C}_5\, [64] = \{ n_0 + n_1\, 5 + n_2\, 5^2 \,:\,\,
n_i = 1, 2, 3, 4 \}\,,   \label{2.1}\end{equation} where $n_i$ are
digits different from zero. This is a finite expansion to the base
$ 5$. It is obvious that $5$ is a prime number and that the set
$\mathcal{C}_5 [64]$ contains $64$ natural numbers. It is
convenient to denote elements of $\mathcal{C}_5 [64]$ by their
digits to the base $5$ in the following way: $ n_0 + n_1\, 5 +
n_2\, 5^2 \, \equiv n_0\, n_1\, n_2$. Here ordering of digits is
the same as in the expansion and it is opposite to the usual one.

Now we are interested in $5$-adic distances between elements of
$\mathcal{C}_5\, [64].$  It is worth recalling  $p$-adic norm
between integers, which is related to the divisibility of integers
by prime numbers. Difference of two integers is again an integer.
$p$-Adic distance between two integers can be understood as a
measure of divisibility  of their difference by $p$ (the more
divisible, the shorter). By definition, $p$-adic norm of an
integer  $m \in \mathbb{Z}$, is $|m|_p = p^{-k}$, where $ k \in
\mathbb{N} \bigcup \{ 0\}$ is degree of divisibility of $m$ by
prime $p$ (i.e. $m = p^k\, m'\,, \,\, p\nmid m'$) and $|0|_p =0.$
This norm is a mapping from $\mathbb{Z}$ into non-negative
rational numbers.  One can easily conclude that $0 \leq |m|_p \leq
1$ for any $m\in \mathbb{Z}$ and any prime $p$.

$p$-Adic distance between two integers $x$ and $y$ is
\begin{equation}
d_p (x\,, y) = |x - y|_p \,.    \label{2.2}
\end{equation}
Since $p$-adic absolute value is ultrametric, the $p$-adic
distance (\ref{2.2}) is also ultrametric, i.e. it satisfies
inequality
\begin{equation}
d_p (x\,, y) \leq\, \mbox{max}\, \{ d_p (x\,, z) \,, d_p (z\,, y)
\} \leq d_p (x\,, z) + d_p (z\,, y) \,, \label{2.3}
\end{equation}
where $x, \, y$ and $z$ are any three integers.

 $5$-Adic distance between two numbers $a, b \in \mathcal{C}_5
\, [64]$ is

\begin{equation} d_5 (a,\, b) = |a_0 + a_1 \, 5 + a_2 \, 5^2 - b_0 - b_1 \, 5
- b_2 \, 5^2 |_5 \,,   \label{2.4} \end{equation}
where $a_i ,\,
b_i \in \{ 1, 2, 3, 4\}$. When $a \neq b$ then $d_5 (a,\, b)$ may
have three different values:
\begin{itemize}
\item $d_5 (a,\, b) = 1$ if $a_0 \neq b_0$, \item $d_5 (a,\, b) =
1/5$ if $a_0 = b_0 $ and $a_1 \neq b_1$, \item $d_5 (a,\, b) =
1/5^2$ if $a_0 = b_0 \,, \,\,a_1 = b_1$ and $a_2 \neq b_2 $.
\end{itemize}
 We
see that the largest $5$-adic distance between numbers is $1$ and
it is maximum $p$-adic distance on $\mathbb{Z}$. The smallest
$5$-adic distance on the  space $\mathcal{C}_5 \, [64]$ is
$5^{-2}$. Note that $5$-adic distance depends only on the first
two digits of different numbers  $a, b \in \mathcal{C}_5 \, [64]$.

Ultrametric space $\mathcal{C}_5 [64]$ can be viewed as 16
quadruplets with respect to the smallest $5$-adic distance, i.e.
quadruplets contain 4 elements and $5$-adic distance between any
two elements within quadruplet is $\frac{1}{25}$. In other words,
within each quadruplet elements have the first two digits equal
and third digits are different.

  With respect to  $2$-adic distance, the above quadruplets may be viewed
 as composed of two doublets: $a = a_0\, a_1\, 1$ and $b = a_0\, a_1\, 3$
 make the first doublet, and
 $c = a_0\, a_1\, 2$ and $d = a_0\, a_1\, 4$ form the second one. $2$-Adic
 distance between codons within each of these doublets is
 $\frac{1}{2}$, i.e.
 \begin{equation}
d_2 (a,\, b) = |(3 -1)\, 5^2|_2 =\frac{1}{2} \,, \, \quad \, d_2
(c,\, d) = |(4 -2)\, 5^2|_2 =\frac{1}{2} \,.   \label{2.12}
 \end{equation}
 By this way ultrametric space $\mathcal{C}_5 [64]$ of 64
elements is arranged into 32 doublets.


Identifying  appropriately nucleotides by digits, we obtain the
corresponding ultrametric structure of the codon space in the
vertebrate mitochondrial genetic code. We take the following
assignments between nucleotides and digits in $\mathcal{C}_5
[64]$: C (cytosine) = 1,\, A (adenine) = 2,\, T (thymine) = U
(uracil) = 3,\, G (guanine) = 4. There is now evident one-to-one
correspondence between codons in three-letter notation and
three-digit $n_0\, n_1\, n_2$ number representation of ultrametric
space $\mathcal{C}_5 [64],$  see Table 1.

The above introduced set $\mathcal{C}_5\, [64]$ endowed by
$p$-adic distance we  call {\it $5$-adic codon space} (or {\it
$5$-adic space of codon states}), because elements of
$\mathcal{C}_5\, [64]$ represent codons (or codon states) denoted
by $n_0 n_1 n_2$.

Let us now explore distances between codons. To this end, it is
useful to look at the Table 1 as a representation of the
$\mathcal{C}_5\, [64]$ codon space. We observe that there are 16
quadruplets such that each of them has the same first two digits.
Hence $5$-adic distance between any two different codons within a
quadruplet is
\begin{align} d_5 (a,\, b) &= |a_0 + a_1 \, 5 + a_2 \, 5^2 - a_0 -
a_1 \, 5 - b_2 \, 5^2 |_5 \nonumber
\\ & = |(a_2 - b_2) \, 5^2|_5 = |(a_2 - b_2)|_5 \,\, | 5^2 |_5 =
5^{-2}\,, \label{2.11} \end{align} because $a_0 = b_0$, $a_1 =
b_1$ and $|a_2 - b_2|_5 = 1$. According to (\ref{2.11}) codons
within every quadruplet are at the smallest $5$-adic distance,
i.e. they are closest comparing to all other codons.

Since codons are composed of three ordered  nucleotides, each of
them is either  purine or  pyrimidine, it is desirable  to
quantify nearness inside purines and pyrimidines, as well as
distinction between elements from these two groups of nucleotides.
This is natural to do by $2$-adic distance. Namely, one can easily
see that $2$-adic distance between pyrimidines  C and U is $d_2
(1, 3) = |3 - 1|_2 = 1/2$ as the distance between purines  A and G
is $d_2 (2, 4) = |4 - 2|_2 = 1/2$. However $2$-adic distance
between C and A or G as well as distance between U and A or G is
$1$ (i.e. maximum).

\begin{table}
   {\begin{tabular}{|l|l|l|l|}
 \hline \ & \ & \ & \\
  111 \, CCC \, Pro &   211 \, ACC \, Thr  &  311 \, UCC \, Ser &  411 \, GCC \, Ala  \\
  112 \, CCA \, Pro &   212 \, ACA \, Thr  &  312 \, UCA \, Ser &  412 \, GCA \, Ala  \\
  113 \, CCU \, Pro &   213 \, ACU \, Thr  &  313 \, UCU \, Ser &  413 \, GCU \, Ala  \\
  114 \, CCG \, Pro &   214 \, ACG \, Thr  &  314 \, UCG \, Ser &  414 \, GCG \, Ala  \\
 \hline \  & \  &  \ & \ \\
  121 \, CAC \, His &   221 \, AAC \, Asn  &  321 \, UAC \, Tyr &  421 \, GAC \, Asp  \\
  122 \, CAA \, Gln &   222 \, AAA \, Lys  &  322 \, UAA \, Ter &  422 \, GAA \, Glu  \\
  123 \, CAU \, His &   223 \, AAU \, Asn  &  323 \, UAU \, Tyr &  423 \, GAU \, Asp  \\
  124 \, CAG \, Gln &   224 \, AAG \, Lys  &  324 \, UAG \, Ter &  424 \, GAG \, Glu  \\
 \hline \  & \  & \  &   \\
  131 \, CUC \, Leu &   231 \, AUC \, Ile  &  331 \, UUC \, Phe &  431 \, GUC \, Val  \\
  132 \, CUA \, Leu &   232 \, AUA \, Met  &  332 \, UUA \, Leu &  432 \, GUA \, Val  \\
  133 \, CUU \, Leu &   233 \, AUU \, Ile  &  333 \, UUU \, Phe &  433 \, GUU \, Val  \\
  134 \, CUG \, Leu &   234 \, AUG \, Met  &  334 \, UUG \, Leu &  434 \, GUG \, Val  \\
 \hline \ & \   & \  &   \\
  141 \, CGC \, Arg &   241 \, AGC \, Ser  &  341 \, UGC \, Cys &  441 \, GGC \, Gly  \\
  142 \, CGA \, Arg &   242 \, AGA \, Ter  &  342 \, UGA \, Trp &  442 \, GGA \, Gly  \\
  143 \, CGU \, Arg &   243 \, AGU \, Ser  &  343 \, UGU \, Cys &  443 \, GGU \, Gly  \\
  144 \, CGG \, Arg &   244 \, AGG \, Ter  &  344 \, UGG \, Trp &  444 \, GGG \, Gly  \\
\hline
\end{tabular}}{} \vskip3mm
{{\bf Table 1.} The $p$-adic vertebrate mitochondrial genetic
code. }
\end{table}

By the above application of $5$-adic and $2$-adic distances to
$\mathcal{C}_5\, [64]$ codon space we have obtained internal
structure of the codon space in the form  of doublets. Just this
$p$-adic structure  of codon space with doublets corresponds to
the mapping which we find in  the vertebrate mitochondrial genetic
code. The other (about 20) known versions of the genetic code in
living systems can be viewed as slight modifications of this
mitochondrial code, presented at the Table 1.

\section{Conclusion}

In this paper I have discussed some aspects of measurements,
$p$-adic numbers and their applications in mathematical physics
and related topics. In particular, I used $p$-adic strings and the
genetic code to demonstrate how $p$-adic modelling works and can
be useful.

It is also emphasized that results of measurements are rational
numbers with Archimedean norm and natural ordering. Thus, $p$-adic
numbers are not results of measurements, but nevertheless they
play important role in $p$-adic mathematical physics and its
applications. In particular, one can expect their further
significant role in description of information sector of the
living systems (see, e. g. \cite{khrennikov}).

\section*{Acknowledgements}
Research presented in this paper is supported by Ministry of
Education and Science of the Republic of Serbia, grant No 174012.

\end{document}